\documentclass[aps,prl,showpacs]{revtex4-1}
\usepackage{ulem}
\usepackage{amsmath}
\usepackage{amsfonts}
\usepackage{amssymb}
\usepackage{fancyhdr}
\usepackage{color}
\usepackage[utf8]{inputenc}
\usepackage{mathtools}
\usepackage{graphicx}
\usepackage{subcaption}
\usepackage{float}
\usepackage{physics}
\usepackage{bbold}
\usepackage{url} 
\usepackage{color}
\usepackage{cancel}
\usepackage{array} 
\usepackage{latexsym}
\usepackage{mathrsfs}
\usepackage{enumitem}
\DeclareGraphicsExtensions{.pdf,.png,.jpg}
\usepackage{array}
\usepackage{verbatim}
\usepackage[colorlinks=true, urlcolor=blue,citecolor=blue,linkcolor=blue]{hyperref}
\begin{document}

\title{Non-classicality of bright GHZ-like radiation of an optical parametric source. } 
\author{Konrad Schlichtholz, Bianka Woloncewicz, Marek \.Zukowski}
\affiliation{International Centre for Theory of Quantum Technologies (ICTQT),
University of Gdansk, 80-308 Gdansk, Poland}
\begin{abstract}
With the emerging possibility to obtain  emissions of triples of GHZ-entangled photons via a direct parametric generation we study here bright emissions of this kind which involve higher order emissions  of two triples, three triples, etc. Such states would constitute a natural generalization of the four mode (two beams plus polarization) squeezed vacuum. We have three beam process of emission generalized bright GHZ states, which a are superpositions of one, two, three, and so on GHZ  entangled triples of photons. We show how to avoid technical difficulties related with straight ahead generalization of the usual description of  parametric down conversion. Using Pade approximation we turn first terms of the non-converging perturbation expansion into elements of conversing series. This allows us to study non-classicality of the new bright generalized  GHZ states.  
\end{abstract}

\maketitle

\section{Introduction}

Multiphoton interferometry is extensively  studied and used in context of  revealing non-classical phenomena \cite{ALLZUKO}. However, in the majority of  such experiments emissions, or rather detections, of fixed numbers of photons is used. Parametric down conversion (PDC) became  robust source of entangled photon pairs \cite{ENTPAIRS}, three and  four photon entangled states \cite{EFFGHZ}, \cite{4BOUW}, \cite{ZUKOHERALD} \cite{4FOTONY}, cluster and Dicke states \cite{HERALDCLUSTER}, \cite{HERALDSTEAM},\cite{SIXPHOTON}. These states find use in  testing  fundamental laws of quantum mechanics and Bell inequalities and demonstration of applications  of  quantum information theory \cite{TELEPORT}, quantum metrology \cite{METRO}, cryptography, communication protocols \cite{PROTO1}, \cite{PROTO2}, imaging \cite{DZIADY} and related topics. 
Thus, nowadays PDC is considered as one of versatile tool to demonstrate non-classicality or quantum  communication protocols, etc., with of quantum optics.

Still ``bright'' states of undefined photon number e.g. bright squeezed states of light  \cite{30BSV}, can also exhibit quantum properties, \cite{BOUW}, \cite{ZUKUMASHA}, and be used to demonstrate e.g. EPR-Bell non-classicality. Such states can be generated in non-linear process of parametric down conversion when we allow for strong pumping  \cite{MASHA}.

An  emblematic example of non-classical light of undefined photon number is  $2\times 2$ mode bright squeezed vacuum generated via type II parametric down conversion which exhibits EPR-like anticorrelations of  Stokes observables for the two beams \cite{MASHAphi}. Its singlet-like form gives invariance of polarization effects under any pair of unitary identical transformations of polarization performed on both subsystems. Thus it is commonly considered as a generalization of the Bell singlet state, as it is sharing a lot of its properties\cite{ROSOLEK}.
 
The following question emerges. As the $2 \times 2$ mode squeezed vacuum can be a generalization of the singlet Bell state, can we have similar analogues for GHZ states, i.e. states that demonstrate quantum features of GHZ state for qubits and simultaneously have undefined photon number? Can they be obtained via  suitable PDC process? 

A parametric process which   produces  photon correlations in three beams via emissions of triples is well-defined quantum optically \cite{GENSQUEEZ} and hence can be achievable in the laboratory. Several experimental attemps to obtain three photon down conversion were performed  successfully \cite{MASHA3FOT}, \cite{TrzeciORDER}, \cite{MEKSYK}. Yet, non-linear crystals are not  the only possible tool used to obtain such states. In \cite{3FOTONY} authors report an observation of three photon parametric down conversion in  a superconducting parametric cavity. As the process is becoming experimentally feasible, it is the hihest time to give to it an effective theoretical description, and see what types of non-classicality can we expect..

It was shown in \cite{IMPOSIBRU} that a straightforward generalization of the usual approximate description of PDC processes  to three  photon emissions is impossible.  The source of problems is the parametric approximation in which emitted photons are treated in quantum optical way, while the pump field is approximated by the classical wave. This approximation works perfectly fine for $2$-photon down conversion but  its generalization to three (or more) photon processes is impossible. One must describe the pump field as quantum one, and in this moment the description becomes much more involved. 

We present  a new approximate  method of how to avoid mathematical difficulties emerging for higher order parametric Hamiltonians. Our approach is a hybrid of theoretical derivation followed  by a numerical approximation method. The nature of the qubic nonlinearity of crystal polarization is such that only three beam emission process is feasible i.e. one pump photon splitting into three down-converted photons. Thus, we show  specific results and figures only for the three beam case. However, we discuss also currently infeasible higher order processes (requiring even higher nonlinearities, and thus most probably out of experimental reach).  

\section{Bright GHZ states}
In the famous EPR paper \cite{EPR} the authors  describe a thought experiment which in their opinion pointed at incompleteness of quantum mechanics. In Bohm's version of EPR experiment a particle of spin $0$ decays into two $\frac{1}{2}$ spin particles in singlet state that are  sent in  opposite directions \cite{BOOOOM}. The particles are correlated in such a way that after performing a spin component measurement on first particle  one can predict with certainty the result of a measurement of the same spin component of second particle, i.e. we are able to predict the result of a  remote identical measurement without performing actual measurement on the other particle. Thus, following EPR, such result must be an ,,element of physical reality''. As ,,elements of reality'' are not present in quantum mechanical description, EPR concluded that quantum theory is not complete \cite{EPR}. Such was the birth of local realism. 

However in 1964 Bell has shown, that it is impossible to construct a local realistic (LHV) model that would explain all possible correlations between two such spins and  would agree with  statistical predictions of quantum mechanics for measurements of arbitrary pairs of spin components. In 1989 Greenberger, Horne and Zellinger (GHZ) showed that for three or four spins one can show directly that the concepts of ,,elements of reality'' is at odds with quantum predictions \cite{GHZ}. 

With emerging bright parametric sources of three beam entanglement one should check to what extend the highly non-classical properties of GHZ states are also shared with their ``bright'' versions.

\subsection{Pitfalls of the parametric approximation (classical pump)}

Here, we shall study technicalities concerning theoretical description of multiphoton GHZ-like  state (bright GHZ) of $n$ beams of light which is a kind of $n$ beam generalization of $2\times 2$ mode squeezed vacuum. We assume that each beam  has two orthogonal polarization modes, but equivalently one can imagine that the consider $n$ pairs of beams, each pair directed to a different observer who is equipped with a Mach-Zehneder interferometer, into which the local beams enter (each via a different entry port). Such an interferometer is capable to perform any $U(2)$ transformation of the pair of modes, and thus it is endowed with powers to show the same type of interference effects as universal polarization beamsplitter. Thus we shall use the ``polarization picture'' throughout just for the simplicity of presentation, but we do not suggest here that the polarization vesion of the experiment would be more feasible (as a matter of fact it seems less feasible, due to a complicated phase matching required for such a case, whereas for two beams per observer situation is much more clearer).

Let us introduce the following notation: 
\begin{align}
&\hat{A_n^\dagger}=\prod_{X=1}^n\hat{a}_X^\dagger, \\
&\hat{B_n^\dagger}=\prod_{X=1}^n\hat{b}_X^\dagger,
\end{align}
where   
$a_X^\dagger$ and $b_X^\dagger$ are creation operators for two orthogonal  polarization modes of $X$-th party's beam.

Parametric approximation in  which the "pump" is described  as a classical field \cite{OPTICSCOM}, \cite{ZUKUMASHA},\cite{ZUKOHERALD} leads to the following Hamiltonian:  
\begin{equation}
\hat H_{n}=\textcolor{black}{\gamma}(\hat A_n^\dagger +\hat B_n^\dagger) +h.c., 
\label{HamGHZ}
\end{equation}
where $\gamma$ is an effective  coupling with the classical pumping field. Parametric approximation is simple, very intuitive and widely used in the description of quantum system interacting  with intense electromagnetic field \cite{ZUKOHERALD}, \cite{ZUKUMASHA}, \cite{30BSV}. 
For $n=2$ the unitary  transformation with Hamiltonian (\ref{HamGHZ}) acting on vacuum state produces a $2\times 2$ mode squeezed vacuum state, with perfect correlations for Stokes observables, which is an analogue of two-qubit   Bell state: $\ket{\Phi^+}$ \cite{MASHAphi}. Still, as any other approaximation, this one also has a range of applicability that requires investigation in every considered case. The approximation is cause no mathematical problems in the case  of two photon  down conversion. Still, it is known that for $n>2$ and expression of the form $\exp{it\gamma H_n}$ is not a well defined unitary transformation (the expansion series does not converge) \cite{IMPOSIBRU} \cite{OPTICSCOM}. Thus, a straightforward generalization for $n>2$ is impossible. Still, one can show that for the pump treated as a (coherent) quantum field, the fully quantum Hamiltonian leads to a well defined evolution. This approach is way more demanding that parametric one. Nevertheless, this is not the only option. The Hamiltonian (\ref{HamGHZ}) can be used with suitable approximation that allows convergence of perturbation series (for example see: \cite{GENSQUEEZ}).

\subsection{Convergence via Pad\'e method}

Our approach is based on two steps. First we expand $e^{iH_{n}t}$ acting on the vacuum state. As said earlier there are problems with convergence. To address this problem, we apply Pad\'e approximants. By combining  the two approximations we get a convergent formula. 
 
Consider the following Hamiltonian: 
\begin{equation}
\label{MINIHAM}
H^A_n= \gamma\hat  A^\dagger_n + h.c.
\end{equation}
For $n =1$ the unitary transformation with  Hamiltonian (\ref{MINIHAM}) produces a coherent state and for $n=2$ two-mode  squeezed vacuum. For $n =3$ the  Hamiltonian (\ref{MINIHAM})  corresponds to  the  Hamiltonian presented in \cite{3FOTONY}. Thus after time $t$ we seem to have: 
\begin{equation}
\label{MULTIFOT0}
\ket{\Sigma^n}=e^{i H^A_nt}\ket{\Omega}=\sum_{k=0}^{\infty}\frac{(i\Gamma)^k}{k!}(\hat A^\dagger_n + \hat A_n)^k\ket{\Omega}, 
\end{equation}
where $\Gamma=\gamma t$ is the amplification gain. But the formula (\ref{MULTIFOT0}) for more than two parties is meaningless! The  series in (\ref{MULTIFOT0}) does not converge i.e. sum of probabilities tends to infinity instead of 1. Thus, (\ref{MULTIFOT0}) is not well defined state. The vacuum state is not an analytical vector for unitary transformation based on Hamiltonian (\ref{MINIHAM})  for  $n>2$, \cite{IMPOSIBRU}. 
However, the expansion of (\ref{MULTIFOT0}) is only a formal description which an approximate form of the Hamiltonian.  We shall introduce an additional, compensatory approximation that allows convergence.

First, 
note that (\ref{MULTIFOT0}) can be put as follows:
\begin{equation}
\ket{\Sigma^n}=
\sum_{k=0}^{\infty}C_k^n(\hat A^\dagger_n)^k\ket{\Omega},
\label{MULTIFOT}
\end{equation}
where $C^k_n$ are coefficients. We show in Appendix that $C^k_n$ can be expanded as follows:
\begin{equation}
C^n_k=\sum_{l=0}^{\infty}\frac{(i\Gamma)^{n+2l}}{(k+2l)!}P^{k,n}_{k+2l},
\label{CNsy}
\end{equation}
where $P^{k,n}_l$  obey the recurrence relation:
$
P^{k,n}_l=P^{k-1,n}_{l-1}+(k+1)^n P^{k+1,n}_{l-1}
 $ Still, the series (\ref{CNsy}), just like (\ref{MULTIFOT0}),  does not converge. Its infinite sequence of the partial sums does not have a finite limit. To impose convergence  we  use a numerical method of Pad\'e approximants \cite{PAD}. 

\subsubsection{Characteristics of n-mode squeezed-like state with Pad\'e approximants and \textcolor{black}{convergence of photon number}}
\label{PAD}
Even if power series does not converge we can still derive an alternative convergent approximation. Note, that the reason of non-convergence is the fact that we threat the pumping field as classical. Thus, we do not have an overall energy conservation in the case of emitted photons. Still, it is obvious that only first dozen-or-so of the expansion terms matter, because the process of emission is of a very low probability. Hence, we shall seek for an approximation that is suitable for such a case.  There are many methods that allow to extract  information from power series outside of its convergence radius. One such method, which is extensively used in numerical calculation, is Pad\'e approximants.

\textcolor{black}{Pad\'e approximants are based on the idea of reformulating power series $\sum c_nx^n$ into a limit of a sequence of ratio of polynomials. Elements of this sequence have the following form:
\begin{equation}
Q_M^N(x)=\frac{\sum_{n=0}^{N}X_nx^n }{\sum_{{m}=0}^{M}Y_{{m}}x^{{m}}}, \label{pppp}
\end{equation}
where $X_n$ and $Y_m$ are such that the  first $(N+M+1)$ terms of the Taylor series expansion of 
$Q_M^N(x)$ match  the first $(N+M+1)$ terms of $\sum c_nx^n$.}

We denote by [N/M] the respective $Q_M^N(x)$. We use diagonal series of approximates  i. e. [N/N] and the highest degree of approximants is [40/40] in order to avoid machine epsilon and \textcolor{black}{other numerical errors}.

Still, we must remember that convergence of coefficients in the Fock space is not sufficient itself. We must also ensure convergence of average  photon number of the  superposition (\ref{MULTIFOT}) - that convergence will determine the range of applicability of  Pade approximants for our expansion. 
To test our method we reconstructed the expansion coefficients for $n =1$ case, i.e. those for a coherent state and the coefficients for $n = 2$ that is for two- mode squeezed vacuum (generated by PDC), see table I below. 

\begin{table}[ht]
	\label{tabel1}
\begin{tabular}{|l|l|l|l|}
\hline
\hline
k  & $n=3$  & $n=2$   & $n=1$               \\ \hline
0  & 0.60   & 0.55    & 0.53                \\ \hline
1  & 0.16   & 0.24    & 0.34                \\ \hline
2  & 0.074  & 0.11    & 0.11                \\ \hline
3  & 0.040  & 0.048   & 0.023               \\ \hline
4  & 0.024  & 0.021   & 0.0037              \\ \hline
5  & 0.016  & 0.0093  & 0.00047             \\ \hline
6  & 0.011  & 0.0041  & 5$\cdot 10^{-5}$    \\ \hline
7  & 0.0087 & 0.0018  & 4.6$\cdot 10^{-6}$  \\ \hline
8  & 0.0066 & 0.0008  & 3.7$\cdot 10^{-7}$  \\ \hline
9  & 0.0052 & 0.00035 & 2.6$\cdot 10^{-8}$  \\ \hline
10 & 0.0042 & 0.00016 & 1.7$\cdot 10^{-09}$ \\ \hline
\end{tabular}
\caption{Probability $p(k)$ of observing $k$ single photons emitted in a coherent state $(n=1)$, photons pairs from two mode-squeezed vacuum $(n=2)$, and triples of photons from three beam radiation $(n=3)$. The range of $k$ is $1,...,10$ and the  calculation is performed for constant value of amplification gain $\Gamma = 0.8$. For $n=2$ obtained values are consistent with theoretical results. Note that probability of vacuum increases with $n$. This is due to the fact that higher order states are generated in processes of a higher degree of non-linearity. As $k$ increases, probability starts to increase with  $n$. The probability of observing $k$ triples of photons for three beam radiation is higher then probability of getting $k$ pairs of  photons for $n=2$ for the same $\Gamma$ starting from $k =4$.} 
\end{table}

\begin{figure}[ht]
	\centering	
	\includegraphics[scale=0.7]{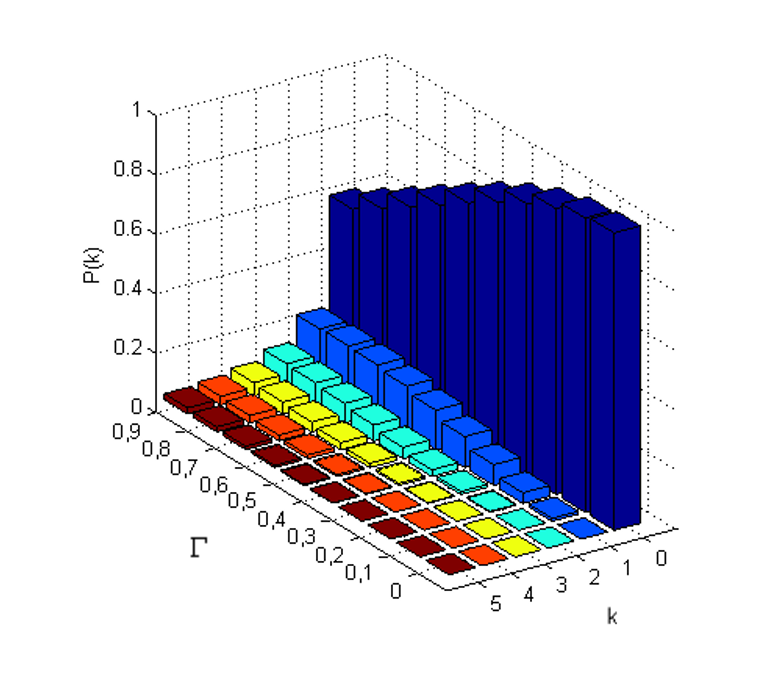} 
	\caption{Probability $p(k)$  of observing $k$ triplets of photons  for  $k =0....5$ in function of  the amplification gain $\Gamma$.
		The probability of vacuum event decrease  when $\Gamma$ increases. Also all probabilities approach zero  when number of triples goes to infinity: $ p(k\rightarrow\infty) \rightarrow 0$. This tendency is typical  also for two mode squeezed vacuum. 
	}
	\label{3dgGam}
\end{figure}

Let us consider the  problem of convergence of the total photon number. The problem was pointed out in \cite{HILZUB}.
We shall define the range of amplification gain for which expectation value of photon number  converges. If
$p(k)$ converges faster then $\sum_{k=1}^{\infty}\frac{1}{k^2}$ from some $k$, the average photon number is  always finite.
  For the realistic case of  $n=3$ the critical applicable amplification gain is around $\Gamma = 0.9$.  Thus, here our approximation breaks down and therefore the results $\Gamma$ approaching $0.9$ are most probably not describing the real situation. 
Fig. \ref{3dgGam} shows  probabilities $p(k)$ of observing $k$ triples of photons  in function of the amplification gain $\Gamma$.

\section{Non-classical properties of 3-party 6-mode Bright GHZ}

Applying the results from previous sections  we are  going to construct a GHZ-like state $\ket{BGHZ}$  which can be generated with  use of Hamiltonian (\ref{HamGHZ}) for $n=3$.
\textcolor{black}
{Since operators $\hat A_3$ and $\hat B_3$  commute we  have  }
\begin{equation}
e^{i\Gamma(\hat A^\dagger_3+\hat A_3+\hat B^\dagger_3+\hat B_3)}=e^{i\Gamma(\hat A^\dagger_3+\hat A_3)}e^{i\Gamma(\hat B^\dagger_3+\hat B_3)}.
\end{equation}
Thus, we can put the state into the following form:
\begin{equation}
\ket{BGHZ}=\sum_{k=0}^\infty\sum_{m=0}^k C^3_{k-m}C^3_m (\hat A^\dagger_3)^{k-m}(\hat B^\dagger_3)^m\ket{\Omega}, 
\label{BGHZ0} 
\end{equation}
where $C^3_Q$ for $Q=k$ or $Q = k-m$ \textcolor{black}{can be obtained with  Pade approximants described in previous sections. From (\ref{BGHZ0}) we can see that the state $\ket{BGHZ}$ is symmetric under change of indices $(k-m\leftrightarrow m)$ i. e. amplitudes of probability for states obtained by action of operators $(\hat A^\dagger_3)^{q}(\hat B^\dagger_3)^p$ and $(\hat A^\dagger_3)^{p}(\hat B^\dagger_3)^q$ on vacuum are equal 
}. 
 
We denote by $\hat T$ the correlation tensor, the  elements of which
are given by: $T_{ijk} = \langle \hat S_i^1 \hat S_j^2 \hat S_k^3\rangle$, where $\hat S_q^X$ is $q$-th normalized Stokes operator  for $X$-th party introduced in \cite{HE}, and rediscovered in \cite{ZUKUBELL} the form which we use here. For a general theory of such quantum Stokes operators see \cite{GENERAL}. These operators for $X$-th party can be represented with photon number operators for respective modes as follows:
\begin{equation}
\langle \hat S_{q}^X \rangle = \langle \hat \Pi^X \frac{(\hat n_{j}^X - \hat n_{{j}_\perp}^X)}{(\hat n_{j}^X + \hat n_{{j}_\perp}^X)} \hat \Pi^X\rangle.
\label{STOKES0}
\end{equation}
In the formula  $j, j_{\perp}$ 
denote a pair of orthogonal  
polarizations  of one of three mutually unbiased 
polarization bases $j = 1,2,3$. Further down 
 we assign index $1$ for polarizations $\{ {45^{\circ}}, {-45^{\circ}}\}$, index $2$ for $\{ {R}, {L}\}$  (circular) and index $3$ for the  $\{ {H}, {V}\}$ basis. The projector $\hat \Pi^X = 1-\ket{\Omega^X}\bra{\Omega^X}$ where $|\Omega^X\rangle$ 
is the vacuum state in $X$-th party, makes the formula well-defined, as it does not allow zero eigenvalues for the denominator. 
The zeroth operator is  $\langle \hat S_0^X \rangle  = \langle \hat \Pi^X\rangle$.     
The normalized quantum optical Stokes operators (\ref{STOKES0}) allow one to straightforwardly introduce Bell inequalities for optical fields, based on photon number observables,  \cite{ZUKUBELL} and \cite{GENERAL}.

One can show (see Appendix)  that non-vanishing elements  of $\hat T$  are 
$T_{111} =t$  and  $ T_{122} = T_{212}= T_{221} = - t$, where $t$ is given by:
\begin{equation}
t=\sum_{k=1}^\infty \sum_{m=0}^k\left((C_{k-m-1}^3)^*(C_{m+1}^3)^*\frac{((k-m)!(m+1)!))^{3}}{k^3}+(C_{m-1}^3)^*(C_{k-m+1}^3)^*\frac{(m!(k-m+1)! ))^{3}}{k^3}\right) C_m^3C_{k-m}^3.\label{tensor}
\end{equation}

Thus, we have the same set of non-vanishing elements of correlation tensor with the same relative  signs, as for three qubit (spin $\frac{1}{2}$) $\ket{GHZ}$ state. This observation  shows that indeed $\ket{BGHZ}$   has the same type of correlations as a three qubit state which is a mixture of "white noise"  and a $\ket{GHZ}$ state.

\subsection{Mermin-GHZ-like Bell inequality violation by   $\ket{BGHZ}$ }
We are going to derive  Mermin-like Bell inequality  \cite{MERMIN}   for three beam optical fields and local measurements of (normalized) Stokes parameters. 
As it was shown in \cite{ZUKUBELL} thus far we do not have a Bell inequality which involves standard Stokes parameters.

Observer $X$  measures  intensity of a light beam using an analyzer of  $j$-th polarization. 
The outcomes, when one tries to introduce local hidden variables the intensities for the two outputs of the analyzer can be written down as $I_j^X(\lambda)$ and $I_{j_{\perp}}^X(\lambda)$. As we want to model the quantum Stokes parameters, their values are natural numbers (they must agree with the eigenvalue spectrum of the number operators used in  (\ref{STOKES0})).
The symbol $\lambda$ denotes hidden variables and their distribution is denoted $\rho(\lambda)$. 

We with the above model for intensities, introduce local hidden variables $S_j^X(\lambda)$ which represent the predetermined values of stokes parameters (\ref{STOKES0}):
\begin{itemize}
	\item{ for $I_{j}^X(\lambda)+I_{j_\bot}^X(\lambda)\neq 0$:
\begin{equation}
S^X_j(\lambda)= 
\frac{I_{j}^X(\lambda)-I_{j_\bot}^X(\lambda)}{I_{j}^X(\lambda)+I_{j_\bot}^X(\lambda)}
\label{CLASSYOBS}
\end{equation}}
\item{
and if $ I_{j}^X(\lambda)+I_{j_\bot}^X(\lambda) = 0$
then
$S^X_j(\lambda) =0$ see \cite{ZUKUBELL}.}
\end{itemize}

Consider the following expression:
\begin{equation}
S_1^1(\lambda) S_1^2(\lambda)S_1^3(\lambda)- S_1^1(\lambda)S_2^2(\lambda)S_2^3(\lambda)- S_2^1(\lambda)S_1^2(\lambda) S_2^3- S_2^1(\lambda) S_2^2(\lambda) S_1^3(\lambda). \label{exbell}
\end{equation}
The values of $S^X_j(\lambda)$ are bounded by $\pm 1$. As (\ref{exbell}) is linear with respect to all $S^X_j(\lambda)$,
we can find extremal values of  (\ref{exbell}) considering only border values i.e. for which  $|S^X_j(\lambda)| =1$. With that we get the bound of (\ref{exbell}) equal to 2. 
The local hidden variables (LHV) averages for terms of (\ref{exbell}) are given by:
\begin{equation}
\langle S_i^1(\lambda)S_j^2(\lambda)S_k^3(\lambda)\rangle_{LHV}=\int d\lambda \rho(\lambda )S_i^1(\lambda)S_j^2(\lambda)S_k^3(\lambda). 
\end{equation}
Thus, the following generalization of Mermin inequality holds: 
\begin{equation}
|\langle S_1^1(\lambda) S_1^2(\lambda) S_1^3(\lambda)- S_1^1(\lambda)S_2^2(\lambda) S_2^3(\lambda)- S_2^1(\lambda) S_1^2(\lambda) S_2^3(\lambda)- S_2^1(\lambda) S_2^2(\lambda) S_1^3(\lambda)\rangle_{LHV}|\leq 2,
\label{BELL}
\end{equation}

In quantum case if we straightforwardly calculate inequality (\ref{BELL})  for $\ket{BGHZ}$ awe see that it is not  violated, due to the  high probability of vacuum events in $\ket{BGHZ}$. To bypass this problem 
we shall modify inequality (\ref{BELL}). This can be done by reformulating observables (\ref{CLASSYOBS}) in such a way that allows us to assign the value $-1$ for the case when no light detection occures (this concept was first introduced in \cite{ZUKUBELL}). The ideas of \cite{NIEZUKUBELL} were our inspiration. The modified hidden values:
\begin{itemize}
	\item{if $I_{j}^X(\lambda)+I_{j_\bot}^X(\lambda)\neq 0$ we have the same approach as for (\ref{CLASSYOBS}): $ 
S^X_{j}(\lambda) \to
S^{X'}_j(\lambda) = S^X_j(\lambda)$, 
}
  \item{but when  $I_{j}^X(\lambda)+I_{j_\bot}^X(\lambda)= 0$ we assign: $
S^{X'}_j(\lambda) = - 1$.} 
\end{itemize}
Still, the bound for reformulated Bell inequality remain the same, because we have: $ -1 \leq {S_j^{X'}}(\lambda) \leq 1$.
Hence, the modified Mermin-like inequality  has the same form as inequality (\ref{BELL}):
\begin{equation}
\label{BELLNONVAC}
|\langle S_1^{1'}(\lambda)S_1^{2'}(\lambda)S_1^{3'}(\lambda)- S_1^{1'}(\lambda)S_2^{2'}(\lambda)S_2^{3'}(\lambda)- S_2^{1'}(\lambda)S_1^{2'}(\lambda)S_2^{3'}(\lambda)- S_2^{1'}(\lambda)S_2^{2'}(\lambda)S_1^{3'}(\lambda)\rangle_{LHV}|\leq 2.
\end{equation} 

In quantum case we reformulate the normalized Stokes operators in the following way:
\begin{equation}
\hat{S_j^X}\rightarrow \hat{S_j^{X'}}=\hat{S_j^X}-\ket{\Omega^X}\bra{\Omega^X}.
\label{REFORMA}
\end{equation}

For $\ket{BGHZ}$ we get:
\begin{equation}
\label{BELLNONVACq}
\langle \hat{S_i^{1'}}\hat{S_j^{2'}}\hat{S_k^{3'}} \rangle_{BGHZ}=\langle \hat{S_i^{1}}\hat{S_j^{2}}\hat{S_k^{3}}\rangle_{BGHZ}-|\braket{\Omega}{BGHZ}|^2, 
\end{equation} 
because the  expectation values of combination of two Stokes operators and one projector into vacuum vanishes, i.e. : $\langle \hat{S_i^1}\hat{S_j^2}\ket{\Omega^3}\bra{\Omega^3}\rangle_{BGHZ}=0$ as well as $\langle \hat{S_i^1}\ket{\Omega^2}\bra{\Omega^2}\ket{\Omega^3}\bra{\Omega^3}\rangle_{BGHZ}=0$, where $i\neq j$ denote different polarization measurements.

\begin{figure}[ht]
\centering
\includegraphics[scale=0.7]{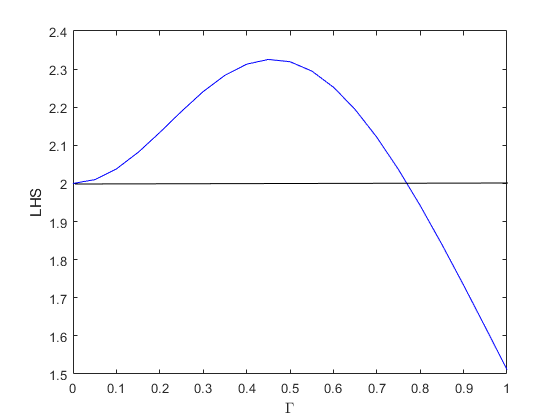} 
\caption{Left-hand side of inequality (\ref{BELLNONVACq}) in function of amplification gain $\Gamma$.
The threshold value of $\Gamma$,  such that for all $\Gamma < \Gamma_{tr}$  inequality (\ref{BELLNONVACq}) is violated is $\Gamma_{tr} =0.77 $.}
\label{belbell}
\end{figure}

Fig.: \ref{belbell} shows the left hand side of inequality (\ref{BELLNONVACq}) in function of the amplification gain. Note that the range of $\Gamma$ for which inequality (\ref{BELLNONVAC}) is violated covers almost all range for $\Gamma$ for which Pad\'e approximation works.   

\subsection{Violation of Mernin-like inequality by $\ket{BGHZ}$ for the case of imperfect detection inefficiency}

We  study here the resistance of the above results with respect to photon losses.
We assume the model of experimental setup in which all photon losses are modeled as inefficient detectors. The standard quantum optical model for that is as follows. The lossy detector is defined as a perfect detector (with efficiency $\eta =1$) with  a beam-splitter  of transitivity  $\sqrt{\eta}$ in front of it with. We assume that in each run of the experiment $k^X$ photons reach  $X$-th observer who has two detectors to detect photons in mutually orthogonal polarization measurements $a$ and $b$. Thus,   $k_a^X+k_b^X=k_X$. However, due to the losses only $\kappa^X_{a_{(b)}}$ counts are registered $(\kappa^X_{a_{(b)}}\leq k^X_{a_{(b)}})$. Probability of outcome  $\kappa^X_i$ for $i =a,b$ including detector efficiency $\eta$ is given by  binomial distribution:
\begin{equation}
p(\kappa^X_i|k_i^X)= {k_i^X\choose \kappa_i^X}\eta^{\kappa_i^X}(1-\eta)^{k_i^X-\kappa_i^X}.
\end{equation}
For simplicity let us consider the expectation value $\langle \hat S'^1_3\hat S'^2_3\hat S'^3_3\rangle$ where  the lower index $3$  stands to  define the measurement basis $\{H,V\}$ (see (\ref{STOKES0})) for $\ket{\phi} = \ket{k^1_H,k^1_V,k^2_H,k^2_V,k^3_H,k^3_V}$. In presence of losses we get:
\begin{align}
\begin{split}
&\expval{\hat S'^1_3\hat S'^2_3\hat S'^3_3}{\phi}=\lim_{\epsilon\rightarrow 0}\prod_{X=1}^3\sum _{\kappa_H^X=0}^{k_H^X}\sum _{\kappa_V^X=0}^{k_V^X}p(\kappa^X_H|k_H^X)p(\kappa^X_V|k_V^X)\Big( \frac{\kappa_H^X-\kappa_V^X}{\kappa_H^X+\kappa_V^X+\epsilon}-\delta_{0,\kappa_H^X+\kappa_V^X}\Big),
\label{BELLNOISE}
\end{split}
\end{align}
where $\delta_{pq}$ describes Kronecker delta. 

In order to calculate other elements of inequality (\ref{BELLNONVAC}) it is enough to apply a unitary transformation that links Stokes operators.

Obviously, the value of  threshold efficiency $\eta_{tr}$ such that for $\eta <  \eta_{tr}$ inequality (\ref{BELLNONVAC}) is not violated,  varies depending of  the amplification gain $\Gamma$. We calculated $\eta_{tr}$ for the range of $\Gamma$ for which  (\ref{BELLNONVAC}) is violated (see Fig. (\ref{BELLNOISE})):
\begin{figure}
\centering
\includegraphics[scale=0.7]{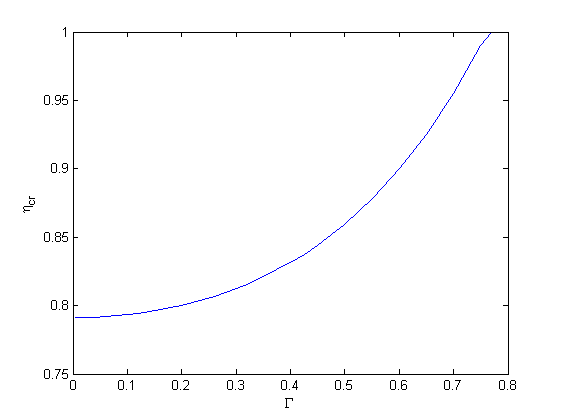} \caption{Threshold efficiency $\eta_{tr}$  in function of amplification gain $\Gamma$. As expected the value of $\eta_{tr}$ increases with $\Gamma$. Note that for small values of amplification gain $\Gamma \to 0$ threshold efficiency $\eta_{tr}=0.79$ and it is conform with $\eta_{tr}$ for qubits given in \cite{EFFGHZ}. That is because for $\Gamma \to 0$  our $\ket{GBHZ}$ becomes effectively a superposition of $\ket{GHZ}$ and vacuum. }
\label{BELLOW}
\end{figure}

\section{Entanglement of $\ket{BGHZ}$}
We present two entanglement indicators for $\ket{BGHZ}$ state. First one is based on an entanglement indicator for $\ket{GHZ}$ for qubits presented in Ref. \cite{TOTH}. The second one is derived from the above Mermin-like Bell  inequality.

The entanglement indicator  for three qubits is given in Ref.  \cite{TOTH}  reads:
\begin{equation}
\label{TOTHW}  
\hat w = \frac{3}{2}\mathbb{1} - \hat \sigma_1^1\hat \sigma_1^2\hat \sigma_1^3   \\- \frac{1}{2} (\hat \sigma^1_3\hat \sigma^2_3 +
\hat \sigma^2_3\hat \sigma^3_3 + \hat \sigma^1_3\hat \sigma^3_3),  
\end{equation}
where $\sigma_k^X$ denotes $k$-th Pauli matrix related with measurement performed on $X$-th party.
Using the isomorphism between Pauli matrices and normalized Stokes operators given e.g. in Ref. \cite{GENERAL}, we straightforwardly obtain  en entanglement indicator for $\ket{BGHZ}$:
\begin{equation}
\label{TOTHSTOKES}
\hat w_1 = \frac{3}{2}\hat S_0^1\hat S_0^2\hat S_0^3 - \hat S_1^1\hat S_1^2\hat S_1^3   - \frac{1}{2} (\hat S^1_3\hat S^2_3\hat S_0^3
+\hat S_0^1\hat S^2_3\hat S^3_3 + \hat S^1_3\hat S_0^2 \hat S^3_3).
\end{equation}
The isomorphism is simply replacement of Pauli operators by normalized Stokes operators, namely $\hat{\sigma}^X_\nu \rightarrow  \hat{S}^X_\nu$, where $\nu=0,1,2,3$.

\begin{figure}
	\includegraphics[scale= 0.80]{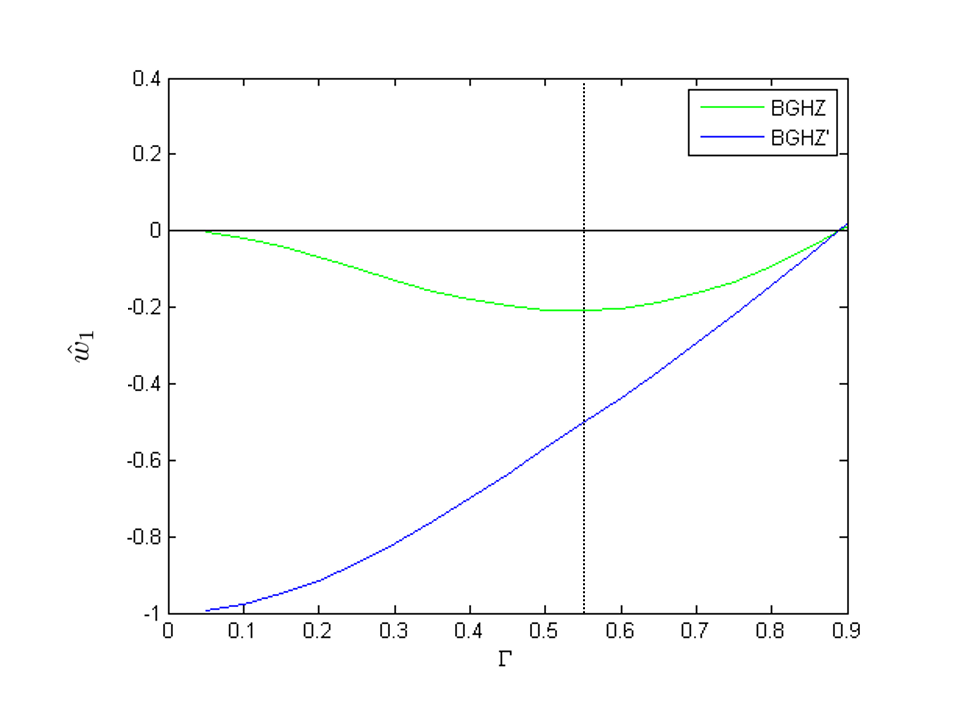}
	\caption{Expectation value of entanglement indicator (\ref{TOTHSTOKES}) for $\ket{BGHZ}$  in function of amplification gain $\Gamma$. The lower blue line is for the state $\rho'$ from which we removed the vacuum cotribution, and renormazeled it, (\ref{PROJPROJ}). This procedure allows one to see better violations for low $\Gamma$, as in the original state in such case dominates the vacuum term. The walue $-1$ points to the theoretical value for the three-qubit entangled GHZ state, and the original entanglement witness (\ref{TOTHW}). This is because for a very small $\Gamma $ the state $\ket{BGHZ}$ is effectively just a superposition of vacuum and a single three photon GHZ emission. Removal of vacuum and renormalization leaves just the GHZ state. }  
	\label{FIGTOTH}
\end{figure}

Using (\ref{TOTHSTOKES}) we can to detect entanglement of  of $\ket{GBHZ}$. Still, the indicator (\ref{TOTHSTOKES}) performs quite weakly e.g.   for small $\Gamma$s it does not detect entanglement. This is due to large amount of vacuum events in $\ket{BGHZ}$. 
To improve detection of entanglement we  use the approach presented in \cite{ZUKUWIESNIAK}. Let us denote density matrix for $\ket{BGHZ}$ as $\hat\rho$. We   remove from $\hat\rho$ where the vacuum contribution in the following way:
\begin{equation}
\hat\rho \to \hat\rho' = \frac{1}{\Tr(\Pi\hat\rho\hat \Pi)}\hat \Pi\hat\rho \hat \Pi, 
\label{PROJPROJ}
\end{equation}
where $\hat \Pi = \hat \Pi^1\hat \Pi^2\hat \Pi^3$. Note that as this is a product of local operations, it does not create new entanglement, and thus the procedure is  admissible.
Figure \ref{FIGTOTH} shows the violation of condition (\ref{TOTHSTOKES}) for $\hat\rho$ and $\hat\rho'$ in function of amplification gain $\Gamma$. 
 

We suggest another entanglement indicator, which  is inspired by the Mermin-like inequality. 
It is well-known that Bel inequalities are entanglement indicators. Still one can improve them in that role, by taking the Bell operators linked with them and calculating, for specific settings, its maximal value for a separable state. This may lead to a lower bound than for local hidden variables (as a separable state can be viewed as a specific local hidden variable model). This allows to create an entanglement indicator (witness) which is more efficient then the initial Bell inequality. 

Consider the following operator:
\begin{equation}
\langle \hat M\rangle = \langle  \hat S_1^1 \hat S_2^2\hat S_2^3+\hat S_2^1\hat S_1^2\hat S_2^3+ \hat S_2^1\hat S_2^2 \hat S_1^3-\hat S_1^1\hat S_1^2\hat S_1^3 \rangle.
\label{BELLWIT}
\end{equation} 
Let us search for its highest value for fully separable states, i.e.
\begin{equation}
\label{ROSEP}
\rho^{1,2,3} = \sum_{\lambda}p_{\lambda}\ketbra{\psi(\lambda)}{\psi(\lambda)},
\end{equation}
where
\begin{equation}
\ket{\psi(\lambda)}_{\lambda} = f^1_\lambda(\hat a^{\dagger})f^2_\lambda(\hat b^{\dagger})f^3_\lambda(\hat c^{\dagger})\ket{\Omega} 
\end{equation}
and $f^1_\lambda(\hat a^{\dagger})$, $f^2_\lambda(\hat b^{\dagger})$ and $f^3_\lambda(\hat c^{\dagger})$ are  functions of powers creation operators for polarization modes corresponding to  optical beams $1,2,3$, of the property that $f^1_\lambda(\hat a^{\dagger})\ket{\Omega}$ is a proper state in the Fock space. 

For every $\lambda$ we have:
\begin{equation}
\langle \hat S_1^1\hat S_1^2\hat S_1^3 \rangle_{\lambda_{sep}} = \langle \hat S_1^1\rangle _{\lambda_{sep}}\langle \hat S_1^2 \rangle_{\lambda_{sep}}\langle\hat S_1^3 \rangle_{\lambda_{sep}}. 
\end{equation}
In order to find the bound it is enough to use the following property of the normalized Stakes operators, see Ref.  \cite{ZUKUWIESNIAK}. If one constructs a 3 dimensional vector $
\langle \vec{\hat S}^X \rangle = (\langle S_1^X\rangle,\langle S_2^X\rangle,\langle S_3^X\rangle)$, one has  $|| 
\langle \vec{\hat S}^X \rangle || \leq 1$, that is  $\langle \hat S_1 \rangle^2 + 
\langle \hat S_2 \rangle^2 +\langle \hat S_3 \rangle^2 \leq 1$. Thus, $\langle \vec{\hat S}^X \rangle$ in a vector in the Bloch ball.

  Note that (\ref{BELLWIT}) involves only two different local measurements, so it is enough to consider only 
 $\langle \hat S_1 \rangle^2 + 
\langle \hat S_2 \rangle^2 \leq 1$. Hence, the relevant Stokes vector of maximal length can  have the following representation:
\begin{equation}
\langle \overrightarrow{\hat S^X} \rangle = (\cos\alpha_X, \sin\alpha_X),
\label{STOKESSINCOS}
\end{equation} 
where $\alpha$ is a certain angle. As the value of operator $\hat{M}$ for pure separable states is proportional to the product of the lengths of the Stokes vectors for each of the beams, using the above representation we search for the maximum of  (\ref{BELLWIT}) by bounding from above the following expression:
\begin{equation}
\begin{multlined}
\cos\alpha_1\sin\alpha_2\sin\alpha_3 + \sin\alpha_1\cos\alpha_2\sin\alpha_3 +\sin\alpha_1\sin\alpha_2\cos\alpha_3 - \cos\alpha_1\cos\alpha_2\cos\alpha_3 \\
=\cos(\alpha_1+\alpha_2+\alpha_3) \leq 1.
\end{multlined}
\end{equation}
Thus we get:
\begin{equation}
-1 \leq \langle \hat{M} \rangle_{sep} \leq 1.
\label{INEQW}
\end{equation}
This bound is by two times smaller than the one for local hidden variable models

Still, inequality (\ref{INEQW}) may be a weak entanglement indicator in case if the state contains a significant vacuum component or admixture. The trick of considering only yhe non-vacuum part of the state, given by (\ref{PROJPROJ})  leads one to a new entanglement indicator (witness)
\begin{equation}
0\leq \langle \hat M + \hat \Pi^1\hat \Pi^2 \hat \Pi^3 \rangle_{sep}  =\langle \hat w_2 \rangle_{sep}.
\label{SUPERCOND}
\end{equation}

\begin{figure}[ht]
	\centering
	\includegraphics[scale=0.7]{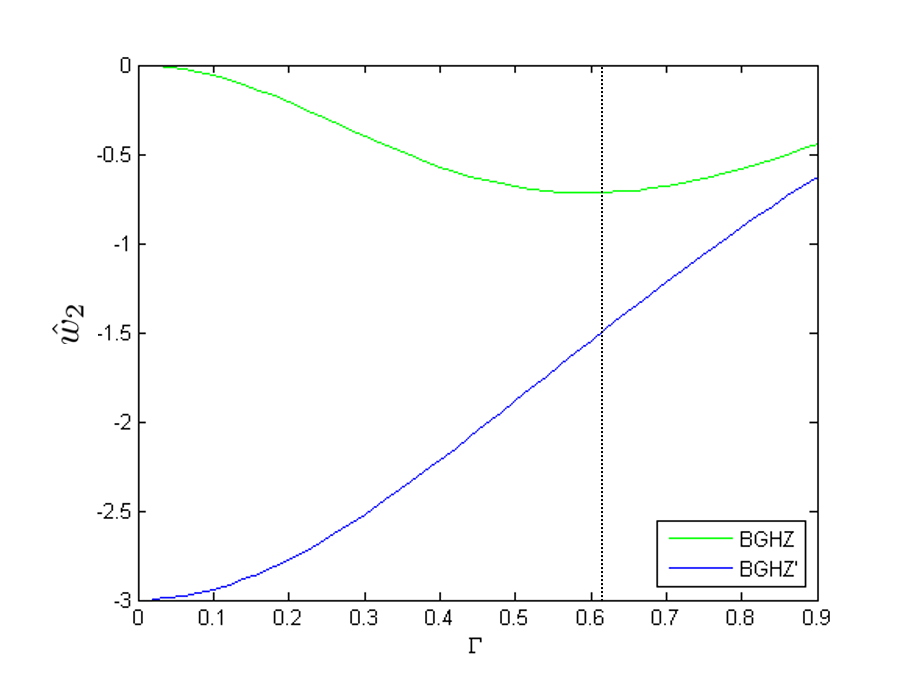} 
	\caption{Comparison of violation of  (\ref{SUPERCOND}) for $\hat \rho$ and $\hat \rho'$ in function of  amplification gain $\Gamma$. 
	See also the caption of Figure 4. Note that in the case of indicator (\ref{SUPERCOND}) we seem to have slightly more robust violations of separability threshold 
	than for (\ref{TOTHSTOKES}) }\label{ENT} 
\end{figure}

For $\ket{BGHZ}$ we get:
\begin{equation}
\expval{\hat w_2}_{BGHZ} = -4t +1 - |C_0^3|^2,
\end{equation}
where $|(C_0^3|^2$ is the probability of vacuum events for all beams.
Fig.: \ref{ENT} shows the violation of separability conditions (\ref{BELLWIT}) and (\ref{SUPERCOND}) in function of amplification gain $\Gamma$. 

When comparing these two entanglement conditions (\ref{TOTHSTOKES}) and (\ref{SUPERCOND}) it seems that (\ref{BELLWIT}), derived from Mermin-like inequality is more efficient than (\ref{TOTHSTOKES}). Comparing Fig.: (\ref{FIGTOTH}) and Fig.: (\ref{ENT}) we see that (\ref{SUPERCOND}) is violated for a broader range of $\Gamma$, and thus it is more robust. For example, $50\%$ of the negative value for $\Gamma \rightarrow 0$ is reached in the case of $\hat{w}_1$ for
 $\Gamma \approx 0.55$, whereas for $\hat{w}_2$ for
 $\Gamma \approx 0.61$, compare the Figures. When analyzing Figure 5 one must have in mind that our approximation breaks down around $\Gamma \approx 0.9$.

\section{Final remarks}   
We have shown that a version of Pad\'e is a candidate for an effective description of the `bright' GHZ states, which cures to some extent the pitfalls of the usual parametric approximation, of the kind that works well for two beam PDC. As such states are interesting and are on the verge of experimental feasibility, the results presented here may be a useful tool for further investigations, and for estimating the influence of the higher order emissions on for example some quantum informational protocols. 
An open question is to find better entanglement indicators, which would be more efficient for higher $\Gamma$'s.

\begin{acknowledgments}
{\it Acknowledgments.}---The work is part of ICTQT IRAP (MAB) project of FNP, co-financed by structural funds of EU.
\end{acknowledgments}

\section{Appendix}
\subsection{\textcolor{black}{Equivalence of representations of $n$ beam multiphoton state: \\ $\ket{\Sigma_n}=\sum_{k=0}^{\infty}\frac{(i\Gamma)^k}{k!}(\hat A^\dagger_n+\hat A_n)^k\ket{\Omega} = \sum_{k=0}^{\infty}C_k^{(n)}(\hat A^\dagger_n)^k\ket{\Omega}$} }
 
We show that formulas (\ref{MULTIFOT0}) and (\ref{MULTIFOT}) are equivalent.

First note that $\hat A_n\ket{\Omega}=0$ and so every therm with operator $A_n$ on the left is equal to zero. 
 Let us now  show that:
\begin{equation}
	\hat A_n^l (\hat A^\dagger_n)^p\ket{\Omega}=\Big(\prod_{j=1}^l(p-j+1)^n\Big)(\hat A^\dagger_n)^{p-l}\ket{\Omega} \label{AAAD3}
	\end{equation}
\textcolor{black}{Let us analyze the action of operator 
$\hat A\hat A^\dagger$  on $n$-beam $k$-photon state $\ket{k_1\cdots k_n}$:
\begin{equation}
\hat A\hat A^\dagger_n\ket{k_1\cdots k_n}
=\Big(\prod_{i=1}^k (k_i+1)\Big)\ket{k_1\cdots k_n},\label{AAAD1}
\end{equation}
because
$
\hat a_X\hat a_X^\dagger\ket{k_1...k_n}=(\hat a_X^\dagger \hat a_X+1)\ket{k_1...k_n}=(\hat n_X+1)\ket{k_1...k_n}=(k_X+1)\ket{k_1...k_n}.$ 
}
Hence,	
\begin{align}
\hat A_n^l (\hat A^\dagger_n)^p\ket{k_1\cdots k_n}=&\prod_{i=1}^n\frac{\sqrt{k_i+p}!}{\sqrt{k_i!}}\hat A_n^{l-1}\hat A_n (\hat A^\dagger_n)\ket{(k_1+p-1)\cdots(k_n+p-1)}\\
=&\Big(\prod_{i=1}^n (k_i+p)\Big)\hat A_n^{l-1}(\hat A^\dagger_n)^{p-1}\ket{k_1\cdots k_n}.
\label{AAAD2}.
\end{align}
Iterating such operation $l$ times we obtain:
\begin{equation}
\hat A_n^l (\hat A^\dagger_n)^p\ket{k_1\cdots k_n}=\prod_{j=1}^l\Big(\prod_{i=1}^n (k_i+p-j+1)\Big) (\hat A^\dagger_n)^{p-l}\ket{k_1\cdots k_n}.\label{italap}
\end{equation}

In case of $\ket{k_1...k_n}=\ket{\Omega}$ relation (\ref{italap}) takes the form of (\ref{AAAD3}).
\textcolor{black}{Thus, $\hat A_n^l (\hat A^\dagger_n)^l$ are effectively only real coefficients, therefore all terms of (\ref{MULTIFOT0}) can be put as $(\hat A^\dagger_n)^l\ket{\Omega}$, for some integer $l$, multiplied by some coefficient. From that we conclude that the formula (\ref{AAAD3}) is correct.}

%
\subsection{Derivation of $C^n_k$ coefficients $C^n_k$}
\textcolor{black}{First, let us notice}, that each coefficient $C^n_k$ is composed of an infinite sum  of  coefficients $P^{k,n}_l$ that stay by $(\hat A_n^{\dagger})^k$ operators linked with  the action of Hamiltonian (\ref{MINIHAM}) on $\ket{\Omega}$. Using the structure of (\ref{MINIHAM}) and state (\ref{MULTIFOT})
we can propose the following form of $C^n_k$:
\begin{equation}
C^n_k = C^n_k=\sum_{l=0}^{\infty}\frac{(i\Gamma)^l}{l!}P^{k,n}_l.
\label{CN_BASE}
\end{equation} 
Now let us define boundary conditions for $P^{k,n}_l$. Note that $P^{k,n}_l = 0$ if $k<0$ or $l<0$. Also analyzing first elements (\ref{MULTIFOT}) we conclude that $P^{k,n}_0 = \delta_{0k}$.  
\textcolor{black}{In the next step, we realize that there are only two ways to obtain term $(\hat A^\dagger_n)^k\ket{\Omega}$ by Hamiltonian acting on $(H^{l-1}_n\ket{\Omega})$:
	\begin{align}
	&\hat A_n(\hat A^\dagger_n)^{k+1}\ket{\Omega}=(k+1)^n(\hat A^\dagger_n)^k\ket{\Omega},  \label{aadn}\\ 
	&\hat A^\dagger_n(\hat A^\dagger_n)^{k-1}\ket{\Omega}=(\hat A^\dagger_n)^k\ket{\Omega}, \label{adadn}
	\end{align}
	where we used identity (\ref{AAAD3}).}

Finally, combining (\ref{aadn}) and  (\ref{adadn})
with the realization that $H^l_n\ket{\Omega}=H_n(H^{l-1}_n\ket{\Omega})$ we deduce the following recurrence pattern for $P^{k,n}_l$:
\begin{equation}
P^{k,n}_l=P^{k-1,n}_{l-1}+(k+1)^n P^{k+1,n}_{l-1} \label{REKURENCJA}.
\end{equation}

Let us now make some remarks on $P^{k,n}_l=0$:
\begin{itemize}
	\item Since $(\hat A^{\dagger}_n + \hat A_n)^k$ acts on 
	$\ket{\Omega}$ 
	(see: \ref{MULTIFOT0}) it appears that  $P^{k,n}_l=0$  if $k>l$. Also 
	$P^{k,n}_k=1$.
	\item  We claim that $P^{k,n}_l=0$ if $(l \mod 2)\neq (k \mod 2)$. The proof  goes via mathematical induction.  One can check easily that $P^{0,n}_1=0$. Let us assume that 
	$P^{k,n}_{k+1}=0$. Using (\ref{REKURENCJA})  we construct $P^{k+1,n}_{k+2}$. We get:
	\begin{equation}
	P^{k+1,n}_{k+2}=P^{k,n}_{k+1}+(k+2)^nP^{k+2,n}_{k+1}=0,
	\end{equation}
\end{itemize}	
\textcolor{black}{
what proves that our assumption is correct for every $n$. Now let us make another induction assumption that $\forall_{k\in N}P^{k,n}_{k+1+2l}=0$ for some $l\in N$.
We calculate  $P^{0,n}_{1+2(l+1)} = P^{1,n}_{2+2l}=0$, due to the assumption. We repeat this reasoning for the case $P^{k,n}_{k+1}$ we conclude that given thesis is correct  and  each second element of infinite sum (\ref{CN_BASE}) vanishes. } 	  	

Consequently, sum (\ref{CN_BASE}) turns to its simplified  form (\ref{CNsy}).

\subsection{Explicit form of $P^{k,n}_l$ coefficients}
\textcolor{black}{Here, we give the explicit form of non-vanishing coefficients 
	$ P^{k,n}_l$ for $l>k$:
	\begin{equation}
	P^{k,n}_l=\sum_{i=1}^{k+1}i^n\sum_{j=1}^{i+1}j^n\cdots \sum_{y=1}^{x+1}y^n, \label{Pexp}
	\end{equation}
	where we have $\frac{l-k}{2}$ sums. The proof is based on mathematical induction and goes as follows.
	Firstly we calculate first few $ P^{k,n}_l$ coefficients that are given in the table (\ref{PnkTab}). We observe that they are obey relation (\ref{Pexp}).}
\begin{table}[ht!]
	\begin{centering}
		\begin{tabular}{|l|l|l|l|l|}
			\hline
			k\textbackslash{}n & 0       & 1       & 2           & 3 \\ \hline
			0                  & 1       & 0       & 0           & 0 \\ \hline
			1                  & 0       & 1       & 0           & 0 \\ \hline
			2                  & 1       & 0 		& 1           & 0 \\ \hline
			3                  & 0 & $1+2^d$       & 0			 & 1 \\ \hline
			4              		&$1+2^d$ &0			&$1+2^d+3^d$ &0 \\ \hline
		\end{tabular}
		\caption{ First few coefficients $P^{k,n}_l$. 
		}\label{PnkTab}
	\end{centering}
\end{table}

Let us assume that relation (\ref{Pexp}) remains valid for $P^{k,n}_{l-1}$. We apply the recursive formula (\ref{REKURENCJA}) for  $l$-th step. We get: 
	\begin{equation}
	P^{k,n}_l=P^{k-1,n}_{l-1}+(k+1)^nP^{k+1,n}_{l-1}.
	\end{equation}
Note that the element $P^{k-1,n}_{l-1}$ is composed from $(l-k)/2$ sums, and $P^{k+1,n}_{l-1}$ has $(l-k-2)/2$ sums.

\begin{equation}
	P^{k,n}_l=\sum_{i=1}^{(k+1)-1}i^n\sum_{j=1}^{i+1}j^n\cdots \sum_{y=1}^{x+1}y^n+(k+1)^n\sum_{j=1}^{(k+1)+1}j^n\cdots \sum_{y=1}^{x+1}y^n.
	\label{ROBO}
	\end{equation}
Also, let us remark that the second term expands the first sum $\sum_{i=1}^{(k+1)-1}$ with the element  corresponding to $i=k+1$. Thus, by merging two terms of (\ref{ROBO}) we get (\ref{Pexp}). 

\section{Derivation of elements of correlation tensor for $\ket{BGHZ_3}$}
\label{TENSSS}
Let us consider state $\ket{BGHZ_3}$ of the following form:
\begin{equation}
\ket{BGHZ_3}=\sum_{k=0}^\infty\sum_{m=0}^k C^3_{k-m}C^3_m (\hat A^\dagger_3)^{k-m}(\hat B^\dagger_3)^m\ket{\Omega}, \label{BGHZr} 
\end{equation} 
Standard Stokes operators for $X$-th beam of light can be written as follows \cite{SIMONBOUV}:
\begin{equation}
\hat \Theta^X_1=\hat b_X^\dagger \hat a_X+\hat a_X^\dagger \hat b_X, \;\;\;\hat \Theta^X_2= i(\hat b_X^\dagger \hat a_X-\hat a_X^\dagger \hat b_X), \;\;\; \hat \Theta^X_3=\hat a_X^\dagger \hat a_X-\hat b_X^\dagger \hat b_X.
\label{STOKES}
\end{equation}
Using formula (\ref{STOKES})  we remind the structure of normalized Stokes operators:
\begin{equation}
\hat S_j^X=\hat \Pi^X \frac{\hat\Theta_j^X}{\hat a_X^\dagger \hat a_X+\hat b_X^\dagger \hat b_X}\hat \Pi^X,
\end{equation}
where $\hat \Pi^X=\hat I^X-\ket{\Omega^X}\bra{\Omega^X}$. 
We are going to calculate elements of correlation tenor for $\ket{BGHZ_3}$. Let us start with  $\langle S_3^1S_3^2S_3^3\rangle$. For simplicity, we use standard Stokes operators $\hat \Theta^X_j$, but our conclusions remain valid for normalized ones. Let standard Stokes operators act on the components of $\ket{BGHZ^3}$ of the following type: $\ket{\phi_{k-m,m}}=\ket{k-m,m}_1\ket{k-m,m}_2\ket{k-m,m}_3$ and  $\ket{\phi_{m,k-m}}$.  We get:
\begin{equation}
\hat \Theta_3^1\hat \Theta_3^2\hat \Theta_3^3\ket{\phi_{k,k-m}}=(k-2m)^3\ket{\phi_{k-m,m}}\label{KETS1}
\end{equation}
and
\begin{equation}
\hat \Theta_3^1\hat \Theta_3^2\hat \Theta_3^3\ket{\phi_{k-m,k}}=(2m-k)^3\ket{\phi_{m,k-m}}.
\label{KETS}
\end{equation}
Thus, due to the symmetry of probability amplitudes in $\ket{BGHZ^3}$ we have the same amplitudes of probability for $\ket{\phi_{m,k-m}}$ and  $\ket{\phi_{k-m,n}}$. Hence, after applying $\bra{BGHZ^3}$  to (\ref{KETS1})and (\ref{KETS}) this two terms cancel out. Obviously, also $S_3^1S_3^2S_3^3\ket{\phi_{k,k}}=0$. Than:$\langle \hat S_3^1\hat S_3^2\hat S_3^3\rangle=0$.

\textcolor{black}{Now, let us make two observations. First, note that  $\hat a_X^\dagger \hat b_X$ and $\hat b_X^\dagger a_X$ flip one photon between modes, without changing total number of photons $X$-th party, meanwhile operators $\hat a_X^\dagger \hat a_X$ and $\hat b_X^\dagger \hat b_X$ does not change number of photons between modes. Also, let us observe that we have to take into consideration  terms of $\langle \hat S_i^1\hat S_j^2\hat S_k^3\rangle$ which are products of operators $\hat A_3$, $\hat A_3^\dagger$, $\hat B_3$, $\hat B_3^\dagger$ because $\ket{BGHZ^3}$ is a superposition of states of the form $(\hat A_3^\dagger)^k(\hat B_3^\dagger)^l\ket{\Omega}$ and so other terms, due to the first observation, acting on $\ket{BGHZ^3}$  will flip photons only in some parties leaving rest of them with unchanged numbers of photons in modes making the state orthogonal to $\ket{BGHZ^3}$.}

\textcolor{black}{Therefore, we can easily conclude that that  all elements of $\hat T$   that contain once or twice $\hat S_3$ operators are equal to zero, because in this case there is no terms which could be nonzero valued. }

\textcolor{black}{Next, we consider element $\langle\hat  S_1^1\hat S_1^2\hat S_1^3\rangle$.} Note that operator  $\hat\Theta_1^1\hat\Theta_1^2\hat\Theta_1^3$ due to the second observation takes effectively the following  form $\hat B^\dagger_3\hat  A_3+\hat A^\dagger_3\hat  B_3$. Hence, we have:
\begin{equation}
\begin{multlined}
(\hat B^\dagger_3\hat  A_3+\hat A^\dagger_3 \hat B_3)\ket{\phi_{k-m,m}} \\=  ((k-m)(m+1))^{3/2}\ket{\phi_{k-1,k-m+1}} 
+((m)(k-m+1))^{3/2}\ket{\phi_{k-m+1,m-1}}
 \label{snm}
\end{multlined}
\end{equation}
and 
\begin{equation}
\begin{multlined}
(\hat B^\dagger_3 \hat A_3+\hat A^\dagger_3 \hat B_3)\ket{\phi_{m,k-m}}= \\ ((m)(k-m+1))^{3/2} \ket{\phi_{m-1,k-m+1}}
+((k-m)(m+1))^{3/2}\ket{\phi_{m+1,k-m-1}}.
\label{smn}
\end{multlined}
\end{equation}
After introducing proper amplitudes of probability and now taking into account the full structure of normalized Stokes operators (with total photon number in the denominator)
we get:
\begin{equation}
\begin{multlined}
m^{3/2}(k-m)^{3/2}C_m^3C_{k-m}^3\bra{BGHZ_3}\hat S_1^1\hat S_1^2\hat S_1^3\ket{\phi_{k,k-m}} \\= \Big((C_{k-m-1}^3)^*(C_{m+1}^3)^*\frac{((k-m)!(m+1)!)^{3}}{k^3} \\
+(C_{m-1}^3)^*(C_{k-m+1}^3)^*\frac{(m!(k-m+1)! )^{3}}{k^3}\Big) C_m^3C_{k-m}^3
\label{F1}
\end{multlined}
\end{equation}
and
\begin{equation}
\begin{multlined}
m^{3/2}(k-m)^{3/2}C_m^3C_{k-m}^3\bra{BGHZ_3}\hat S_1^1\hat S_1^2\hat S_1^3\ket{\phi_{k-m,k}} \\= \Big((C_{k-m-1}^3)^*(C_{m+1}^3)^*\frac{((k-m)!(m+1)!)^{3}}{k^3} \\
+(C_{m-1}^3)^*(C_{k-m+1}^3)^*\frac{(m!(k-m+1)! )^{3}}{k^3}\Big) C_m^3C_{k-m}^3.
\label{F2}
\end{multlined}
\end{equation}
Note that formulas (\ref{F1}) and (\ref{F2}) are equal. Finally, we obtain 
\begin{align}
\langle \hat S_1^1\hat S_1^2\hat S_1^3\rangle=&\sum_{n=1}^\infty \sum_{m=0}^n\Big((C_{k-m-1}^3)^*(C_{m+1}^3)^*\frac{((k-m)!(m+1)!)^{3}}{k^3} \\
&+(C_{m-1}^3)^*(C_{k-m+1}^3)^*\frac{(m!(k-m+1)! )^{3}}{k^3}\Big) C_m^3C_{k-m}^3. \label{s1}
\end{align}

Let us now consider element that contain one operator $\hat S_1$ and two operators  $\hat S_2$.  Analogously to  the  reasoning above, the product of such operators turns to  $i^2(\hat B^\dagger_3 \hat A_3+(-1)^2\hat A^\dagger_3\hat B_3)=-(\hat B^\dagger_3 \hat A_3+\hat A^\dagger_3\hat B_3)$. Thus, this elements have the same absolute  value as  
 (\ref{s1}).

Next, we investigate element $\langle\hat  S_2^1\hat S_2^2\hat S_2^3\rangle$. Again, this element effectively gives $i^3(\hat B^\dagger_3 \hat A_3+(-1)^3\hat A^\dagger_3\hat B_3)=-i(\hat B^\dagger_3 \hat A_3-\hat A^\dagger_3\hat B_3)$. It is also easy to notice that the expectation value of this operator is zero (respective terms  cancel out) 

Also, element with two operators $\hat S_1$  and one $\hat S_2$  can be written as follows: $i(\hat B^\dagger_3\hat  A_3-\hat A^\dagger_3\hat B_3)$. Thus this element  is equal to $-\langle \hat S_2^1\hat S_2^2\hat S_2^3\rangle = 0$

\section{Unitary transformations that link Stokes operators in context of mutually umbiased basis.}

According to formula (\ref{STOKES0}) one can express Stokes operators as the ratio of difference and sum of photon number operators related with perpendicular polarization vectors of a given polarization basis. These bases are mutually unbiased i.e. the square of modulus of scalar product between vectors, each from different base, is equal to the inverse of dimension of the basis e.g $|\braket{H}{R}|^2 = \frac{1}{2}$. Two vectors from different polarization basis are linked with $2 \times 2$  unitary transformation.

Let 
$\hat a_j^\dagger (m)$ be a
 creation operator of $j$-th polarization mode related with  $m$-th Stokes operator $\hat S_m$ such that:
 $\hat S_m  =  \frac{\sum_{j=0}^1(-1)^j\hat a^{\dagger}_j(m)\hat a_j(m)}{\sum_{j=0}^1\hat a^{\dagger}_j(m)\hat a_j(m)}$. 
The indices $j=0$ and $j=1$ stand for two perpendicular polarization vectors of a given basis. 
Consider $2 \times 2$ unitary matrices  related with unitary transformations leading from one polarization basis to another. Using them we can represent creation operators for $m=1,2$ through creation operators for $m=3$ as follows:
\begin{equation}
\label{TRANS}
\hat a_j^\dagger(m)  = \sum_{s=0}^1 U_{js}(m)\hat a^{\dagger}_s(3),
\end{equation}

A transformation of type of (\ref{TRANS}) can be performed also on Stokes operators, see e.g. \cite{GENERAL}. 
Applying this reasoning  to the formula (\ref{BELLNOISE}) we can recover all needed terms from (\ref{BELLNONVACq}).


\end{document}